# The missing link in nanomaterials governance: industrial dynamics and downstream policies

Ismael Rafols[1,2], Patrick van Zwanenberg[1]
Molly Morgan[1], Paul Nightingale[1] and Adrian Smith[1]

[1] SPRU – Science & Technology Policy Research, Freeman Centre
University of Sussex, Brighton BN1 9QE, UK

[2] Technology Policy & Assessment Center, School of Public Policy,
Georgia Institute of Technology, Atlanta, GA 30332, USA.

9th August 2009

**Abstract**

In this article we explore the analytical and policy implications of widening the governance of nanomaterials from the focus on risk regulation to a broader focus on the governance of innovation. To do this, we have analysed the impact of industrial activities on nanotechnology governance, while previous studies have concentrated on risk appraisal, public perceptions, public engagement, regulatory frameworks and related policies. We argue that the specific characteristics of the industrial dynamics of nanomaterials – *flexibility in applications* and *distributed innovation* - have important implications for innovation governance as they limit and enable different interventions to attempt to shape technology. Flexibility and distributedness exacerbate the difficulties of directly controlling or shaping the directions of nanomaterials innovation. In particular, the potential for public policy leverage is hindered by the bottom-up nature of governance resulting from the multiplicity of innovation sites. Under these conditions, we argue that the framing of policies for nanomaterials governance needs to be broadened. The prevailing emphasis on policy initiatives upstream, while commendable, should be complemented with broader downstream policies, such as renewable energy procurement or regulations in housing, in order to modulate technological development towards socially desirable goals.

**Keywords:** nanotechnology; manufactured nanomaterials; innovation governance; industrial dynamics; enabling technology; technology assessment; public engagement.

## 1. Introduction: from risk governance to innovation governance

The development of nanotechnology has been accompanied by a high level of social awareness about the need to address public concern over health, environmental safety and the social implications of the new technologies. This increased awareness of the influence of public perceptions on innovation (following experience with genetically modified organisms) has led to policy efforts to promote the development of socially acceptable technologies that do not compromise human health or environmental safety. These efforts have resulted in a variety of expert committees, public engagement experiments and technology assessment exercises aimed at improving nanotechnology governance (e.g. RS/RAE 2004; Renn and Roco 2006a; SCENIHR 2007; Gavelin et al. 2008; RCEP 2008).

Three main conclusions have repeatedly emerged from these efforts. First, 'nanotechnology' is a palette of disparate technologies at the nano-scale and does not necessarily constitute a useful category to discuss regulation or technology governance (Doubleday 2007; RCEP 2008, p.12; Rip 2009). Second, there is major uncertainty and ignorance regarding the safety of many manufactured nanomaterials and, since for some nanomaterials 'there is a plausible basis for concern that harmful effects might arise', a precautionary approach is recommended (RCEP 2008, p. 27; Hansen et al. 2008). Third, the discussion of the risks and benefits posed by nanotechnologies concerns the direction and control of the innovation process and thus should be framed in terms of *innovation governance* rather than *risk governance* (EC 2007). Risk governance is preoccupied with minimising the risks of harmful effects of nanotechnologies – hence it is a back-end response to innovation. Innovation governance, on the other hand, is aimed at purposefully influencing the technological choices made so that innovation is directed towards socially agreed goals, such as health, well-being, social justice and environmental sustainability. Note that here we adopt a broad conceptualisation of governance, as 'all structuring of action and interaction that has some authority and/or legitimacy' whether or not these actions or interactions are intentionally aimed at governing or regulating (Rip 2009, p. 2).

The first and second conclusions (disparity of nanotechnologies and ignorance on risks) have been widely discussed and we will not dwell further on them, although their uptake in policy is limited. In this paper we focus on the third aspect, the move from risk to innovation governance, which has received far less discussion and has yet to be translated successfully into policy. The UK's Royal Commission on Environmental Pollution (RCEP) report on novel materials (RCEP 2008), in which the authors of this paper were indirectly involved,[1] exemplifies that despite awareness of a broader perspective, the shift in policy beyond risk to innovation governance remains elusive. This RCEP report makes forceful proposals to address risk governance, spelling out specific measures, such as advocating for a partial revision of REACH[2], mandatory reporting of nanomaterials manufacturing and usage and an early warning system. In some passages it does also discuss the challenge of governing the direction of emergent technologies, but when it comes to recommendations, it resorts to rather general exhortations for 'an adaptive governance regime capable of monitoring technologies and materials as they are developed and incorporated into processes and products' (RCEP 2008, p. 8) and a recognition of 'the importance of continual "social intelligence" gathering and the provision of ongoing opportunities for public and expert reflection and debate' (RCEP 2008, p. 79). Its proposals serve mainly to reinforce the framing of policies in terms of singular risks and overlook the wider innovation dynamics. It does not address what a governance regime should look like, nor how, aside from preventing harm, policies could influence the direction of technical change and the ultimate uses of the technology.[3]

---

[1] We produced a scoping study for the RCEP on *Nanomaterials Innovation Systems: Their Structure, Dynamics and Regulation* (see Nightingale et al. 2008), part of which was used in the official RCEP report (RCEP, 2008). Here we present a summary of the main insights of the scoping study. Readers are referred to Chapter 3 of it, Nightingale et al. (2008), for a presentation of the theoretical framework used, based on the innovation systems literature and evolutionary perspectives on science and technology.
[2] REACH is the new European Union regulation concerning the Registration, Evaluation, Authorisation and restriction of CHemicals.

Nevertheless, if not explicit in the framing of governance, in practice one can observe a growing concern for the direction of nanotechnology innovation. On the one hand, at a normative level, the discourse on responsible development of nanotechnology 'implies a commitment to develop and use technology to help meet the most pressing human and societal needs, while making every reasonable effort to anticipate and mitigate adverse implications or unintended consequences' (National Research Council 2006, p. 73). This remit is clearly difficult since neither the social benefits nor the risks are well understood.[4] On the other hand, in purely instrumental terms, application of nanomaterials is shifting in search for public acceptability. This is because public attitudes to the risks posed by nanomaterials (as in biotechnology) differ widely depending on their application, for instance, in food or health, and on the perceptions of who reaps the benefits and who bears the costs (RCEP 2008, p. 72; Stirling 2008). Not surprisingly under these circumstances, *green* nanotechnologies have become the new and acceptable side to 'nano' in contributing to the solution of the world's most pressing problems such as clean water and climate change (Jones 2007; Schmidt 2007; Rebuffat 2008). Hence, for both normative and instrumental reasons, nanomaterials development is under pressure to follow particular technological pathways in preference to others. This means that, in reality, a *de facto* governance of technological choices is in place, orchestrated by networks of disparate actors and conflicting aspirations, interests and norms (Rip 2009). The political concern is that the particular interests of powerful actors in governance networks may result in certain technological developments being favoured at the expense of more 'plural priorities' (Stirling, 2009). However, it is unclear how the governance of an emergent technology might be made more explicit, transparent and democratic. And it is not obvious whether nanomaterials innovations can (or should) be purposefully directed.

In the succeeding sections of this paper we present an analysis of innovation governance for the specific case of manufactured nanomaterials,[5] with the aim of making more explicit the sites of technological choice and exploring the scope for public policy intervention. Many previous studies of nanomaterials governance focus on risk appraisal (Renn and Roco 2006b; Hansen et al. 2008), public perceptions and deliberation (Kahan et al. 2009; Pidgeon et al., 2009), soft and hard regulatory frameworks (Lee and Jose 2008; Stokes 2009) and the entanglement of these with policy (Macnaghten et al. 2005; Doubleday 2007). A few studies analyse the impact of industrial activity on risk governance, for example, in relation to life cycle assessment (Bauer et al. 2007; Meyer et al. 2009), but very few discuss the impact of industrial dynamics on the direction of technical change, i.e. in innovation governance (Robinson, 2009). We believe that this is a crucial missing link in our understanding of nanomaterials governance.

---

[3] Similarly, the recent UK Government response to the RCEP report frames governance in terms of understanding and management of risks. Innovation and 'the capture of economic and social benefits' are regarded as separate issues, which are left mainly to industries and their exercise of corporate social responsibility (UK Government 2009, p. 24). In section 4 we argue that these issues are (a key) part of *de facto* governance (Rip 2009).
[4] Given current levels of uncertainty and ignorance about the risks posed by some nanomaterials, this statement would suggest the adoption of a (potentially contentious) precautionary approach: the development of some nanomaterials would be justified only if their applications are aimed at socially relevant goals.
[5] Hereafter the discussion is presented exclusively in terms of manufactured nanomaterials, a specific subset of nanotechnologies. They are among the 'first' generation nanotechnologies, which have reached or are about to reach commercialisation and thus pose the most urgent threat. This does not mean that innovation governance analysis for other nanotechnologies, e.g. 'active nanotechnologies' ('second' generation), is not relevant. But it also does not follow that our analysis can be generalised. See Doubleday (2007) for a discussion on the appropriate scales for framings of nanotechnologies.

In this article we argue that the specific characteristics of the industrial dynamics of nanomaterials have important implications for innovation governance as they limit and enable different interventions that attempt to shape technology. The article is organised as follows. Section 2 reviews the key concepts that underlie the view that the directions of innovation can be purposefully influenced. Section 3 argues that, since nanomaterials are enabling technologies, they can be used as 'products for process innovation' for a variety of applications in myriad commercial sectors – and their function and use (and meaning) can be re-interpreted at different points in diverging value chains. It follows that the industrial dynamics of nanomaterials is characterised by *flexibility of applications* and *distributed innovation*. These characteristics have two implications. The first, discussed in Section 4, is that nanomaterials development is intractable in terms of direct control, that is, towards *specific* technological targets; however, there is room for emergent, *de facto* governance to be 'modulated' (Rip 2009). The second implication, discussed in Section 5, is that policy interventions to shape technological development are needed in the early technology stage, but not only upstream (Wilsdon and Willis 2004) and midstream (Fisher et al. 2006; Joly and Rip 2008) but also downstream, closer to the end-user. Section 6 provides a summary and concludes the article.

2. *Shaping the direction of innovation*

Attempts to influence the direction of innovation in novel materials are beset by the so-called Collingridge (1992) dilemma, which holds that we cannot control technologies once they are well developed because they become entrenched, thereby making them susceptible to various forms of 'lock-in' and developmental momentum in a particular direction. Once material investments, infrastructures, institutional commitments, routines and user habits develop around a particular technology, the development trajectory for the whole technological system becomes established and extremely difficult and costly to reverse or redirect (Dosi 1982; Arthur 1989; Williams 2000; Shove 2003). A classic example is the QWERTY keyboard which was intentionally designed to be suboptimal to slow down mechanical typewriters, but has persisted in the age of computers (David 1985).

Many have recognised this dilemma and the associated problems affecting innovation and technology development, and a variety of policy approaches have been proposed to try to address this problem of lock-in. Most focus on broadening the scope of societal considerations and stake-holder participation in the processes governing technology before irreversible commitments. Such approaches typically focus on avoiding socially undesirable technological configurations by garnering a wider spectrum of knowledge (see review in Fisher et al. 2006). Practical manifestations of this 'opening up' can be seen in the evolution of technology assessment – initially a technocratic exercise– into 'Constructive Technology Assessment' (CTA) variants, including a broader range of participants and issues (Rip et al. 1995), and 'Real-Time Technology Assessment' (RTTA) variants that emphasise the interactive and reflexive efforts of stakeholders (Guston and Sarewitz 2002). In the last years, strategies of engagement 'upstream' and public deliberation have been proposed, to influence technology in its earliest stages, before lock-in to certain technological trajectories can occur (Wilsdon and Willis 2004; Wilsdon et al. 2005; Macnaghten et al. 2005; Pidgeon et al. 2008). More recently, the discourse on 'responsible development of nanotechnology', with its associated (non-binding) codes of conduct, appears to have emerged as the main normative frameworks for industry (EC 2008; Anon 2008; BASF 2009).

A common theme in all these efforts is that anticipation of future effects should be integrated into the promotional aspects of technology development. Proponents of this view argue that the design of technological development should be a wider, more interactive process that includes a variety of societal actors as well as technical experts. The effect of opening up the design process will be that designers', users', citizens' and policy makers' ideas and values are articulated at an early stage, and should enable valuable contributions to this process and the opening of new areas for innovation.

While these direct and explicit efforts of technology assessment and public engagement are a very valuable contribution to innovation governance, our analysis of innovation dynamics suggests that they are not sufficient to direct innovation (Nightingale et al. 2008).

## 3. *Industrial dynamics: flexibility in applications and distributed innovation*

The key issue regarding industrial dynamics is that manufactured nanomaterials are not consumer products that are sold to end-users, but 'capital' products that are incorporated into other products manufactured by secondary firms in a variety of industries. Products containing nanomaterials can be intermediary products (biosensors) or end-user products (e.g. nano-silver socks) whose performance is improved by the specific properties of the nanomaterials. In other words, nanomaterials are enabling technologies that can be understood as 'products for process innovation'. This has two important implications for their industrial dynamics (and governance) in terms of their *flexibility of application* and the *distributed nature of innovation* (see the map innovation sites in Figure 1).

*Flexibility of application* means that one type of nanomaterial can be used in a variety of applications that can benefit from the same special electronic, optical, catalytic, chemical or physical properties of the material (Aitken et al., 2006, p. 3003).[6] For example, research by Dr Saif Haque of Imperial College London, focuses on charge transfer dynamics in nanostructured molecular materials with the goal of developing solar cells. However, his better understanding of the properties of nanocrystalline $TiO_2$ prompted Haque to branch his research into chemical sensors (e.g. for security checks in airports) and light-emitting displays (e.g. for consumer electronics) –which have completely different social implications (Nightingale et al., 2008, p. 63). Flexibility of application, by definition, is characteristic of

---

[6] Interestingly, the flexibility is two way: since the same specific and special properties are provided by different nanomaterials, choice exists among the materials to be selected to fulfil a given function. For example, carbon nanotubes can be used for applications in drug delivery, photovoltaics and sensors (among others). But for the specific application of photovoltaics, one can use carbon nanotubes, metal oxides nanocrystals or quantum dots. This results in a highly dense matrix of Functions x Nanomaterials, as shown by Aitken et al. (2006, p. 3003).

so-called enabling or general purpose technologies (Youtie et al., 2008). As we will see, one of the key consequences of flexibility is that the 'final outcomes of innovation are not realized until the stages of final implementation/configuration', and may differ extraordinarily from the initial technological visions or promises (Williams, 2006, p. 238)

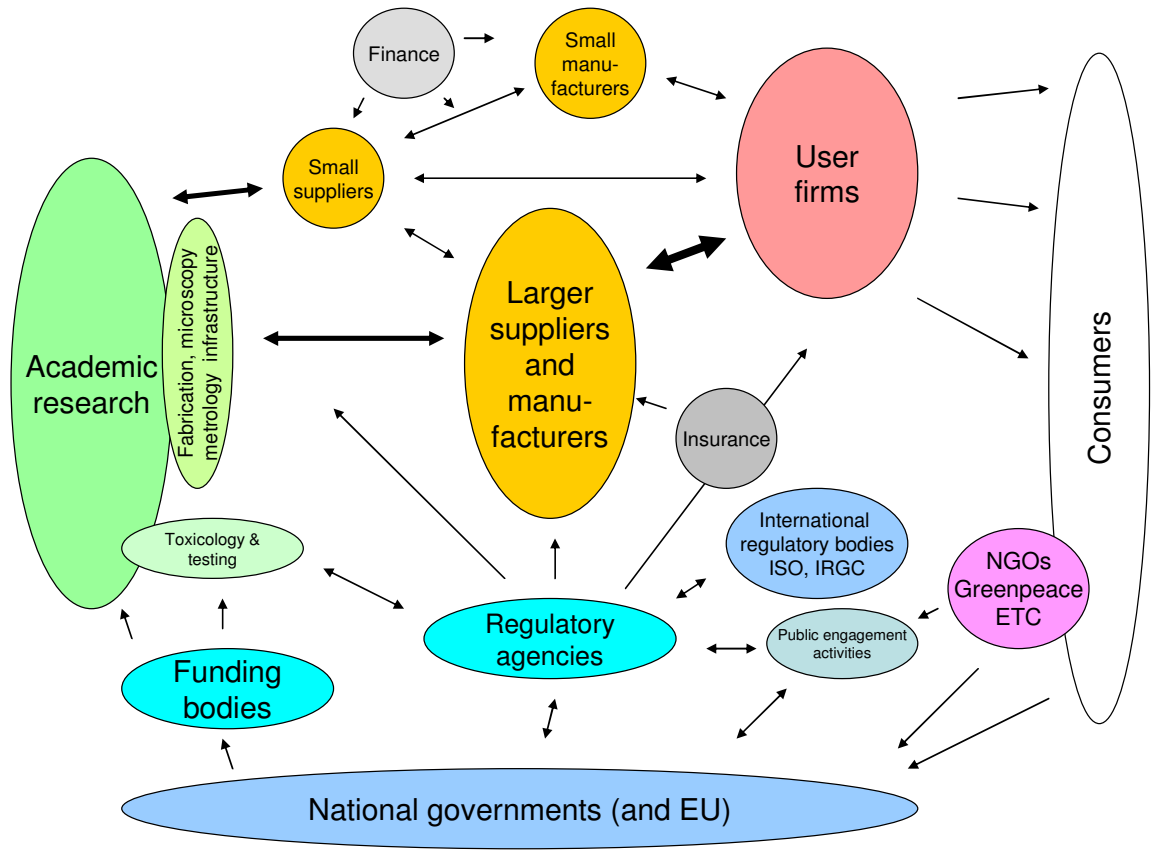

Figure 1. Generic actors and main linkages in nanomaterials innovation and governance networks.

The *distributed nature of innovation* results from the long, branching value chains into which nanomaterials are incorporated as 'products for process innovation'. For example, TiO$_2$ nanoparticles, developed in a collaboration between a university spin-off and a transnational chemical corporation, are bought by manufacturers of ceramic tiles or glass (because they provide self-cleaning properties), who sell their products to construction companies in multiple countries, who sell their buildings to real estate agents, who sell them on to households, who hire people (or do it themselves) to drill into the tiles or windows, remove them and generally redecorate their new homes. In these long industrial networks, innovation occurs at different stages along the value chain[7] – innovation results not only from the synthesis of new nanomaterials (e.g. single wall carbon nanotubes - SWCNTs), but also from new production processes (e.g. use of new catalysts for SWCNTs production), new incorporations of well-known nanomaterials in existing products (e.g. SWCNTs in paper or plastics), new methods of incorporating nanomaterials into products (e.g. ink jet printing of SWCNT), or - more rarely - new end-user products made possible by the properties of

---

[7] Robinson (2009, pp.7-8) proposes a richer description of the value chain in terms of an 'innovation chain' that constitutes a 'mosaic of arenas of innovation and selection'.

nanomaterials (e.g. fully recyclable electronic newspapers enabled by the SWCNT printable memory).[8]

In each of the branching 'junctures' of the value-chain, a given nanomaterial (such as SWCNTs) is subject to multiple pressures and constraints that push, pull and shape its development in certain directions. These influences include the scientific and technical paradigms and routines that frame researchers' thinking (Dosi, 1982); dedicated infrastructures that make substitution with alternatives difficult (Jacobsson and Johnson, 2000); practices that enjoy greater economies of scale and positive network externalities (Arthur, 1989); prevailing social attitudes (Shove, 2003); dominant policies, legal frameworks and professional association lobbying (Walker, 2000); hegemonic discourses that inform socially acceptable performance criteria.[9] Hence, at every stage in the value chain - up-, mid- and downstream, these various factors lead to specific *commitments* that might result in different types of lock-in: to the particular nanomaterials used;[10] to the type of energy or labour required for their production infrastructures; to the possibilities for the product to be recycled; to the types of applications favoured; and so on.

## *4. Accepting lack of control: de facto governance*

Given these conditions of distributed innovation and flexibility in applications, is it actually possible to govern the development of a technology? Some technologies have emerged under relatively controlled environments, where the number of innovating, manufacturing and commercialising companies and other stakeholders is comparatively small and easy to identify. In these cases, one can imagine the possibility of purposefully directing innovation towards specific outcomes via policy interventions that anticipate alternative applications. This might be typical of relatively bounded systems, such as specific energy (e.g. fuel cells) or sanitation (e.g. membrane) technologies, which have well-defined societal functions, or even in generic applications very specifically framed (e.g. 'lab-on-a-chip for cell analysis', Robinson and Propp, 2008). However, in the case of nanomaterials, with an overwhelming multiplicity of innovation sites and actors (distributed innovation) and wide diversity of outcome options and societal functions (flexibility of application), our knowledge is too incomplete and the web of interactions too complex to assume that efforts to 'control' the directions of technology will succeed in driving innovation towards the desired options.

A first response to the perceived intractability of nanomaterials to governance is to reduce the incompleteness of our knowledge. There have been numerous calls for a step increase in the funding for medical and environmental toxicology, and regulatory frameworks (RS/RAE 2004; RCEP 2008; Stokes 2009) in a bid to address the lack of knowledge related to risk issues.[11] Efforts have also been made to increase our understanding of public perceptions, regulations, preferred technological choices and the policy discourses shaping them, often from a risk governance perspective (Kahan et al. 2009; Pidgeon et al. 2009; Stokes 2009; Doubleday 2007). However, with the exception of a few studies in the area of environment, little work has been done on industrial dynamics or nanomaterials value chains (Schmit 2008; Meyer et al. 2009, Robinson, 2009).[12] This lack of research, compounded by the secrecy in nanomaterials trade between manufacturing and user firms, has resulted in a worrying dearth of information on the scale and scope of nanomaterials applications (Maynard and Rejeski 2009). For example, the only public database of nanotech-enabled products is based on information from firms' web sites (PEN 2009) and estimates by private consultants on nanomaterials markets differ by orders of magnitude (Meyer et al. 2008). Hence, more empirical research on the industrial dynamics and life cycle assessment of nanomaterials is needed urgently given the failure of self-reporting schemes (DEFRA 2008) and reluctance (so-far) of most countries to opt for mandatory reporting (UK Government 2009, pp. 19-22).

---

[8] This example is based on media coverage of the report by Song et al. (2008) on inkjet printing using SWCNTs.
[9] E.g., Doubleday (2007) and Robinson (2009) argue that there is lock-in in the policy discourse: the framing of nanomaterials governance is focused narrowly on health, environment and safety issues.
[10] E.g. recent investments in large plants for the production of specific nanomaterials, such as $TiO_2$ or CNTs, may already be leading to lock-in.
[11] Although these same voices acknowledge that even with massive toxicology research, reaching some scientific understanding will take decades – certainly much longer than the time for commercialisation.

However, even if our knowledge of nanomaterials risks, benefits, uses and social preferences were massively boosted, the would-be governance actors would still find the task of controlling and purposefully making choices aimed at specific outcomes overwhelming. This is because, in this case, the combined properties of distributed innovation and flexibility exacerbate the multiplicity of actors, complementary technologies and linkages involved in developing or using nanomaterials. Historical examples of other emergent technologies displaying similar properties, such as chemicals, information and communication technology (ICT) and biotechnology,[13] provide vivid illustrations of the sheer unpredictability of social outcomes, and challenge the assumption that collective action can purposefully shape technological directions (Williams 2006). Yet accepting the incapacity to exert *control* over the *specific* outcomes of nanomaterials, does not negate the possibility of *modulating* their development more broadly towards *generic* goals. However, it does entail adopting a different understanding of what governance constitutes, and where and how it operates, which has policy implications.

In stating that the development of nanomaterials depends on the interactions among multiple actors, and is not amenable to purposeful top-down control, we are implying that the governance of this system is bottom-up and distributed. Arie Rip (2009) introduced the notion of *de facto* governance to describe such a situation. This notion offers two key insights. First, that governance is in place even in the absence of intentional efforts to regulate, control or shape the technology because the existing socio-technical networks both enable and constrain future technological developments. Second, that policies aimed at shaping technologies have to emerge from and act within a given configuration of *de facto* governance, that is, that 'attempts at regulation can be located as interventions in emerging *de facto* governance, and will depend on it for their effectiveness' (Rip 2009, p.1 ).

A variety of activities, such as the discourse on responsible development of nanotechnology and its associated corporate codes, expert committees in the International Organization for Standards, OECD working parties, etc., have been presented as examples of the emergence of a distributed governance for nanomaterials (RCEP 2008; Rip 2009). If we broaden the scope from risk governance to innovation governance – given our interest in technological choice - and adopt a *de facto* governance perspective, it follows logically that not just soft law and associated social arenas, but also a variety of structuring institutions and sites need to be taken into account. A map of the actors playing roles in innovation helps to identify these sites (see Figure 1 in this piece, and Figure 2 in Robinson 2009. p.10). This makes it possible to begin to see how 'distributed innovation and distributed governance merge into one another' (Rip 2008, p. 18). For governments willing to foster certain social applications it is easier to try to act in those areas such as public research, funding and regulatory agencies that are under their direct control.[14] However, it should be emphasised that the core of these innovation networks is configured by the various industrial actors, which consequently play a central role in *de facto* governance. In summary, a *de facto* governance perspective highlights the crucial role played not just by research and development (R&D), and firms manufacturing nanomaterials, but also by the firms making nano-enabled products and other actors such as the insurers, retailers and end-users. This is why the industrial dynamics is such an important site for setting directionality. But, given this strong bottom-up and diffuse nature of governance, what is the room left for policy action to influence technological choices?

---

[12] Research has been conducted on the industrial dynamics of nanotechnology in sociology and economics, but it focuses on the conditions for 'success' for public researchers, firms or clusters on the basis of bibliometric and patent analyses. In our view, this body of work has produced only limited insights relevant to considerations of directionality in innovation.

[13] Note that all these technologies have the characteristics of general purpose technologies, i.e., pervasiveness, innovation spawning, and scope for improvement (Youtie et al., 2008).

## 5. Policies to shape direction: There is plenty of room 'downstream'

We argue that there is room for policy action to shape the directions of nanomaterials innovation, but that the scope of policies needs to be broadened in two respects. First, governance considerations should widen from the current focus upstream to downstream. Second, promotion and prevention instruments should be interwoven in the policy framework.

*Upstream vs downstream policies*
The distributed nature of both innovation and *de facto* governance challenges the view that policies aiming at the shaping of technologies should focus on interventions 'upstream'. In the UK, ambitions to steer nanotechnology innovation in socially-desired directions in the most recent years have been reflected in the widespread support for public engagement 'upstream', in the R&D process (Wilsdon and Willis, 2004; Macnaghten et al., 2005; Pidgeon et al., 2008). The proponents of public engagement (at least the academic social scientists) initially used the term 'upstream' to refer to the 'early stages' of a technology, that is, before the establishment of irreversible commitments, at "a point where research trajectories are still open and undetermined" (Wilsdon et al., 2005, p. 38). However, the notions of 'upstream' and 'downstream' have not always been clearly defined[15] and are frequently conflated in policy practice with the use of upstream in the terminology of industrial value chain analysis (which pre-dated the 'engagement' usage). Within the value-chain terminology downstream refers to those stages closer to the end-user (left-hand side of Figure 1), and upstream to those that lie far from the consumer (right-hand side of Figure 1). This is the meaning we adopt here – reserving 'early stage' to refer to the phase before technological lock-in.

Within the so-called linear model of innovation, which assumes a directional flow of knowledge from science to technology to society, it was easy for non-experts to conflate 'upstream' and 'early-stage' as referring to the same phase, that is, the phase when the technologies are initially developed in the R&D process. However, the linear model has been widely discredited by many historical studies – most technologies are developed and shaped throughout the value chaing, in interactions downstream (e.g. between users and manufacturing firms), as well as midstream (between firms and finance) and 'usptream' (between firms and researchers). Indeed, the description in Section 3 of how innovation occurs at numerous loci in the value chain, in no predetermined order, implies that that the concepts of 'early stage' and 'upstream' should not be confused. The directions of nanomaterials and nano-enabled-products are shaped at the up-, mid-, and downstream sites – and policy intervention (and deliberative practices) may be required 'early-stage' - but in *all* these sites in order to modulate *de facto* governance.

---

[14] E.g., since 2008 the UK Engineering and Physical Science Research Council (EPSRC) has had programmes (Grand Challenges) devoted to the application of nanotechnologies in Energy, Healthcare and the Environment.
[15] One exception is Fisher et al. (2006), for whom 'upstream' is associated with research policy, 'midstream' with R&D practice and 'downstream' with commercial and/or social applications and regulation: 'Upstream decisions may be characterized as determining what research to authorize, midstream decisions as determining how to implement R&D agendas, and downstream decisions as determining whether to adopt developed technologies' (Fisher, 2006, p. 490). Note that this use is slightly different from ours.

*Promotional vs prevention policies*
The governance of technologies traditionally has been split into promotional policies that support research, mainly upstream, and regulation policies aimed at avoiding harm for the end-user, that is downstream (Fisher et al. 2005). A move from risk to innovation governance clearly demands that we overcome the antagonistic distinction between promotion and prevention. Promotion of certain technological pathways prevents the development of others. Prevention via regulation of given technologies indirectly promotes alternatives. The erosion of this distinction is already underway particularly in the area of risk. Under the aegis of the precautionary approach enshrined in EU legislation, risk governance has broadened risk appraisal processes to a wider variety of sites and social actors, including end-users (at the most downstream) as well as researchers and policy makers (upstream). Calls for early warning systems respond to this logic (RCEP, 2008). The analysis developed in the previous two sections suggests that a reciprocal move is also necessary: in order to influence direction, technology promotional policies are needed upstream but also mid- and downstream.

An analysis of the UK's promotional policies for nanomaterials reveals a strong focus on upstream, that is, the left-hand side of Figure 1: funding for basic and applied research; investment for shared facilities and infrastructures; and grants for university-industry collaboration. Yet we found no policies targeted at shaping direction or intervening downstream, where value-chains and markets unfold. This finding suggests there is room for promotional policies intervention downstream. However, the associated unpredictable dynamics of distributed innovation means that many of the policies downstream cannot be geared to the achievement of *specific* outcomes for nanomaterials. Instead policies downstream need to address *generic* frameworks that will have a modulating effect on the development of nanomaterials. This is because policies mid- and downstream (and sometimes upstream), need to perform certain social functions (transportation, energy provision, health), without being specific about the technologies that can provide them.

For example, let us explore the policies that could have an effect on the development of nanomaterials for photovoltaic technologies for sustainable development in commercial sectors such as energy or housing. Given the distributed nature of innovation, the policy measures span both sides of Figure 1 –some are specific to nanomaterials, some are specific to energy or housing, some are generic - either up- or downstream. Policies specific to nanomaterials, mainly up- and midstream, would include metrology equipment, toxicology studies, clearing-houses to facilitate awareness and knowledge flows among universities, manufacturing firms and potential user-firm – which, as argued above, constitute the current policy focus. Also, and equally relevant, would be policies specific to renewable technologies, such as:
- targeted research funding for both public and private R&D;

- the conduct of extended foresight or technology assessment exercises;
- grants to invest in the production capacity required to produce photovoltaics;
- public procurement (e.g. street lighting or government buildings) to encourage the formation of markets;
- clearing-houses to facilitate awareness and knowledge flows between manufacturing firms and potential user-firms;
- targeted regulations, such as a requirement for the incorporation of renewable energies into certain kinds of future housing stock.

In addition, it would be important for these policies to be aligned to generic sustainability policies and pressures within the 'user' sectors (far right-hand side of Figure 1), specific to the energy or housing sectors. This would include, for example, general regulatory environments that encourage investment in renewable technologies, or fiscal policies designed to provide the same incentives to individual households wishing to invest in photovoltaics (Kemp and Pontoglio, 2008).

## 6. Conclusions

This article has explored the analytical and policy implications of moving beyond a policy focus on the governance of risks of manufactured nanomaterials to a focus on the governance of innovation. To do this we analysed the industrial dynamics of nanomaterials, while previous studies on nanotechnology governance had concentrated on risk appraisal, public perceptions, public engagement, regulatory frameworks and related policies. We have argued that nanomaterials are 'products for process innovation', that is enabling technologies that can be used to improve product performance for a variety of applications in myriad commercial sectors in diverging value chains. The ensuing characteristics of the industrial dynamics of nanomaterials – *flexibility in application* and *distributed innovation* – have two important implications for our understanding of how public policy might affect innovation governance. First, since nanomaterials governance has a strong bottom-up nature and is generally intractable in terms of direct control, policy efforts should emphasise the modulation of technologies towards desirable social functions rather than the achievement of specific technologies. Second, in order to succeed, policies aimed at such modulation need to be implemented at a variety of socio-economic sites. This means that policy initiatives upstream make valuable contributions to innovation governance, but they are not sufficient to direct innovation. They need to be complemented by downstream policies such a procurement or targeted regulations. In summary, the analysis of industrial dynamics suggests that a broader understanding of innovation and a richer palette of policies are required if the public policy is to move beyond preventing risk to shaping the social outcomes of nanomaterials development.

**Acknowledgements**
This article is based on research carried during a scoping study for the Royal Commission on Environmental Pollution carried out in 2008 (see Nightingale et al. 2008).